\documentclass{PoS}

\usepackage{graphicx}% Include figure files
\usepackage{subfigure}

\title{Parton Distribution Function from the Hadronic Tensor on the Lattice}

\ShortTitle{Parton Distribution Function}

\author{\speaker{Keh-Fei Liu}\\

        Dept. of Physics and Astronomy,University of Kentucky, Lexington, KY 40506, USA\\
        E-mail: \email{liu@pa.uky.edu}}

\abstract{ \hspace{0.8cm} The path-integral formulation of  the hadronic tensor $W_{\mu\nu}$ of deep inelastic scattering
is reviewed. It is shown that there are 3 gauge invariant 
and topologically distinct contributions. 
The separation of the connected sea partons from those of  the disconnected sea can be achieved with a 
combination of the global fit of the parton distribution function (PDF),  the semi-inclusive DIS  data on the strange PDF 
and the lattice calculation of the ratio of the strange to $u/d$ momentum fraction in the disconnected insertion. 

 \hspace{0.8cm}     We shall discuss numerical issues associated with lattice calculation of the hadronic tensor involving
 a four-point function, such as large hadron momenta and improved
maximum entropy method to obtain the spectral density from the hadronic tensor in Euclidean time.

\hspace{0.8cm}
 We also draw a comparison between the large momentum approach to the parton distribution function (PDF) and
 the hadronic tensor approach.}

\FullConference{The 33rd International Symposium on Lattice Field Theory\\
		14 -18 July 2015\\
		Kobe International Conference Center, Kobe, Japan*}

\begin{document}

\section{Hadronic tensor in path-integral formalism}

The deep inelastic scattering of a muon on a nucleon involves the hadronic
tensor which, being an inclusive reaction, includes all intermediate states
\begin{equation}   \label{w}
W_{\mu\nu}(q^2, \nu) = \frac{1}{2} \sum_n \int \prod_{i =1}^n \left[\frac{d^3 p_i}{(2\pi)^3 2E_{pi}}\right]  \langle N|J_{\mu}(0)|n\rangle
\langle n|J_{\nu}(0) | N\rangle_{spin\,\, ave.}(2\pi)^3 \delta^4 (p_n - p - q) . 
\end{equation}

Since deep inelastic scattering measures the absorptive part of the  
Compton scattering, it is the imaginary part of the forward amplitude and
can be expressed as the current-current correlation function in the nucleon, 
i.e.
\begin{equation}  \label{wcc}
W_{\mu\nu}(q^2, \nu) = \frac{1}{\pi} {\rm Im} T_{\mu\nu}(q^2, \nu)
= \langle N| \int \frac{d^4x}{4\pi}  e^{ i q \cdot x} J_{\mu}(x)
J_{\nu}(0) | N\rangle_{spin\,\, ave.}.
\end{equation} 

It has been shown~\cite{Liu:1993cv,Liu:1998um,Liu:1999ak,Aglietti:1998mz,Detmold:2005gg} that the hadronic tensor 
$W_{\mu\nu}(q^2, \nu)$ can be obtained from the Euclidean path-integral
formalism. 
% where the various parton dynamical degrees of freedom are
%readily and explicitly revealed.
In this case, one considers the ratio of the four-point function  \\
\mbox{$\langle \chi_N(\vec{p},t) \int \frac{d^3x}{4\pi} 
e^{- i \vec{q}\cdot  \vec{x}} J_{\nu}(\vec{x},t_2) J_{\mu}(0,t_1)
\chi_N(\vec{p}, t_0)\rangle$} and the two-point function
\mbox{$\langle \chi_N(\vec{p},t) \chi_N(\vec{p},t_0)\rangle$},
where $\chi_N(\vec{p},t)$ is an interpolation
field for the nucleon with momentum $p$ at Euclidean time $t$.
 
As both $t - t_2 \gg 1/\Delta E_p$ and $t_1 - t_0  \gg 1/\Delta E_p$, where
$\Delta E_p$ is the energy gap between the nucleon energy $E_p$ and the next
excitation (i.e. the threshold of a nucleon and a pion in the $p$-wave),
the intermediate state contributions from the interpolation fields will be dominated by
the nucleon with the Euclidean propagator $e^{-E_p (t-t_0)}$.
From the three-point and two-point functions on the lattice
\begin{eqnarray}  \label{3_pt}
G_{pWp}^{\alpha\beta}& =& \sum_{\vec{x_f}} e^{-i \vec{p}\cdot\vec{x_f}}
 \left\langle\chi_N^{\alpha}(\vec{x_f},t) \sum_{\vec{x}} \frac{e^{-i \vec{q}\cdot \vec{x}}}{4\pi} 
 J_{\mu}(\vec{x},t_2)\,J_{\nu}(0,t_1)  \sum_{\vec{x_0}} e^{i \vec{p}\cdot\vec{x_0}}\,\overline{\chi}_N^{\beta}(\vec{x_0}, t_0)\right\rangle , \\
 G_{pp}^{\alpha\beta} &=&  \sum_{\vec{x_f}} e^{-i \vec{p}\cdot\vec{x_f}}\left\langle \chi_N^{\alpha}(\vec{x_f},t) \,
 \overline{\chi}_N^{\beta}(\vec{x_0} = 0, t_0)\right\rangle, 
 \end{eqnarray}
we define
\begin{eqnarray}  \label{wmunu_tilde}
\widetilde{W}_{\mu\nu}(\vec{q},\vec{p},\tau) &=&
 \frac{E_p }{m_N} \frac{{\rm Tr} (\Gamma_e G_{pWp})}{{\rm Tr} (\Gamma_e G_{pp})}
 \begin{array}{|l} \\  \\  t -t_2 \gg 1/\Delta E_p, \, t_1 - t_0 \gg 1/\Delta E_p \end{array} \nonumber \\
    &=& \frac{E_p }{m_N}\frac{\frac{|Z|^2m_N (E_p + m_N)}{ E_p^2}e^{-E_p(t-t_0)}<N|\sum_{\vec{x}} \frac{e^{-i \vec{q}\cdot \vec{x}}}{4\pi}
  J_{\mu}(\vec{x},t_2) J_{\nu}(0,t_1)|N>}
{\frac{|Z|^2 (E_p +m_N)}{E_p} e^{-E_p(t-t_0)}} \nonumber \\
  &=& <N|\sum_{\vec{x}} \frac{e^{i \vec{q}\cdot \vec{x}}}{4\pi} e^{-i\vec{q}\cdot \vec{x}}
J_{\mu}(\vec{x},\tau) J_{\nu}(0,0)|N>,
\end{eqnarray}
where $\tau = t_2 - t_1$,   $Z$ is the transition matrix element
$\langle 0|\chi_N|N\rangle$, and $\Gamma_e = \frac{1 + \gamma_4}{2}$ is the unpolarized projection to the positive parity nucleon state. 
Inserting intermediate states, 
$\widetilde{W}_{\mu\nu}(\vec{q}^{\,2},\tau)$ becomes
\begin{equation}   \label{wtilde}
\widetilde{W}_{\mu\nu}(\vec{q}^{\,2},\tau)
= \frac{1}{4 \pi}\sum_n \left(\frac{2 m_N}{2 E_n}\right) \delta_{\vec{p}+\vec{q}, \vec{p_n}}\langle N(p)|J_{\mu}(0)|n\rangle
\langle n|J_{\nu}(0) | N(p)\rangle_{spin\,\, ave.} e^{- (E_n - E_p) \tau}.
\end{equation}
Formally, to recover the delta function $\delta(E_n - E_p + \nu)$ in Eq. (\ref{w}) in the continuum formalism, one 
can carry out the inverse Laplace transform with $\tau$ being treated as a dimensionful continuous variable
\begin{equation}  \label{wmunu} 
W_{\mu\nu}(q^2,\nu) = \frac{1}{2m_Ni} \int_{c-i \infty}^{c+i \infty} d\tau\,
e^{\nu\tau} \widetilde{W}_{\mu\nu}(\vec{q}^{\,2}, \tau),
\end{equation} 
with $c > 0$. This is basically doing the anti-Wick rotation back to the 
Minkowski space. 
We will discuss the numerical lattice approach to this conversion from Euclidean space to Minkowski space later. 
%However, this is not practical since Eq.~(\ref{wmunu})  needs data on imaginary $\tau$ 
%where there is no lattice data. Formally, 

\section{Parton degrees of freedom}

    In addressing the origin of the Gottfried sum rule violation, it is shown~\cite{Liu:1993cv,Liu:1998um,Liu:1999ak} that
the contributions to the four-point function of the Euclidean path-integral formulation of 
the hadronic tensor $\widetilde{W}_{\mu\nu}(\vec{q}^{\,2}, \tau)$ in Eq. (\ref{wtilde})  can be classified according 
to different topologies of the quark 
paths between the source and the sink of the proton. Fig. 1(a) and 1(b) represent connected insertions (C.I.) of the
currents.  Here the quark fields from the interpolators $\chi_N$ contract
with the currents such that the quark lines flow continuously from $t =
0$ to $t =t$. Fig. 1(c), on the other hand, represents a disconnected
insertion (D.I.) where the quark fields from $J_{\mu}$ and $J_{\nu}$
self-contract and, as a consequence, the quark loop is disconnected from the quark paths between
the proton source and sink. Here, ``disconnected'' refers only to the 
quark lines. Of course, quarks dive in the background of the gauge field and all quark paths are ultimately
connected through the gluon field.

\begin{figure}[htbp] \label{hadonic_tensor}
\centering
\subfigure[]
{{\includegraphics[width=0.3\hsize]{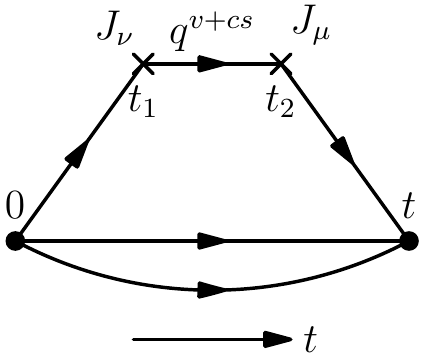}}
  \label{val+CS}}
\subfigure[]
{{\includegraphics[width=0.3\hsize]{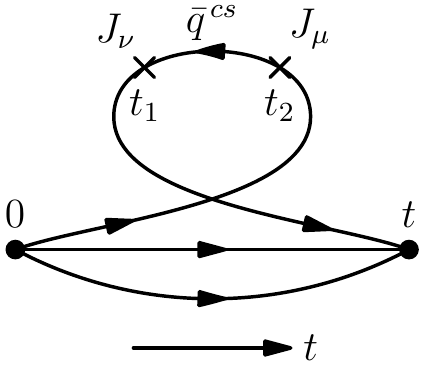}}
  \label{CS}}
%\vspace*{-0.6cm}
\subfigure[]
{{\includegraphics[width=0.3\hsize]{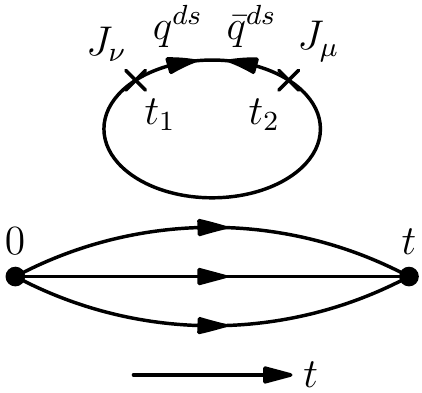}}
 \label{DS}}
\caption{Three gauge invariant and topologically distinct diagrams in the Euclidean-path integral
formulation of the nucleon hadronic tensor in the large momentum frame. In between the currents
at $t_1$ and $t_2$, the parton degrees of freedom are
  (a) the valence and connected sea (CS) partons $q^{v+cs}$, (b) the CS anti-partons $\bar{q}^{cs}$. Only $u$ and
$d$ are present in (a) and (b) for the nucleon hadronic tensor. (c) the disconnected sea (DS) partons $q^{ds}$ and
anti-partons $\bar{q}^{ds}$ with $q = u, d, s,$ and $c$.}
\end{figure}

We first note that Fig.~\ref{CS}, where the quarks propagate backward in time between $t_1$ and $t_2$
corresponds to the connected sea (CS) anti-partons $\bar{u}^{cs}$ and $\bar{d}^{cs}$, since the quark lines are connected
to the nucleon interpolation fields at $t=0$ and $t =t$. By the same token,
Fig.~\ref{val+CS} gives the valence and CS partons $u^{v + cs}$ and  $d^{v + cs}$, 
and the valence is defined as $u^v(d^v) \equiv u^{v + cs}(d^{v + cs}) - \bar{u}^{cs}(\bar{d}^{cs})$
and $u^{cs}(x) \equiv \bar{u}^{cs}(x)$.
On the other hand, Fig.~\ref{DS} gives the disconnected sea (DS) $q^{ds}$ and $\bar{q}^{ds}$ with $\{q = u,d,s,c\}$.  We see that 
while $u$ and $d$ have both CS and DS, strange and charm have only DS.

It is clear from these diagrams that there are two sources of the sea partons, one is CS and the other
is DS. In the isospin limit where $\bar{u}^{ds}(x) = \bar{d}^{ds}(x)$, the DS do not contribute to the 
Gottfried sum rule (GSR) violation which reveals that $\int_0^1 dx [\bar{u}(x) - \bar{d}(x)]  < 0$ from DIS 
experiments. The isospin symmetry breaking due to the $u$ and $d$ 
mass difference should be of the order of \mbox{$(m_d - m_u)/m_N$} and cannot explain the large violation of GSR.
Rather, the majority of the violation should come from the CS. 

As far as the small-$x$ behavior is concerned, there is only reggeon exchange for the flavor non-singlet valence 
and CS, so the small-$x$ behavior for the valence and CS partons is \\
\mbox{$q^{v+cs}(x),\, \bar{q}^{cs}(x) {}_{\stackrel{\longrightarrow}{x \rightarrow 0}} \propto x^{-1/2}.$}
On the other hand, there is flavor-singlet pomeron exchange in addition to the reggeon exchange for the
singlet DS partons, thus its small $x$ behavior is more singular, i.e.
\mbox{$q^{ds}(x),\, \bar{q}^{ds}(x) {}_{\stackrel{\longrightarrow}{x \rightarrow 0}} \propto x^{-1}.$}

In the global fittings of parton distribution function (PDF), the CS is not separated from the DS and it had
been implicitly assumed that all the anti-partons are from the DS. That's why the GSR violation came as
a surprise and the fitting has accommodated the $\bar{u}(x) - \bar{d}(x)$ difference from experiment. However,
it is still mostly assumed that the $\bar{u}(x) + \bar{d}(x)$ has the same $x$ dependence as that of 
$s(x) + \bar{s}(x)$. As we discussed above, $\bar{u}(x) + \bar{d}(x) = \bar{u}^{cs}(x) + \bar{d}^{cs}(x) 
+\bar{u}^{ds}(x) + \bar{d}^{ds}(x) $ have both the CS and DS partons and they have different small $x$ behaviors.
This is in contrast to $s(x) + \bar{s}(x)$ where there are only DS partons. Combining HERMES data on the strangeness 
parton distribution~\cite{hermes08}, the CT10 global fitting of the $\bar{u}(x) + \bar{d}(x)$ distributions~\cite{CT10}, 
and the lattice result of the moment ratio of the strange to $u/d$ in the disconnected insertion, i.e. 
$\langle x\rangle_{s+\bar{s}}/\langle x\rangle_{u+\bar{u}}({\rm DI})$~\cite{doi08}, it is demonstrated ~\cite{Liu:2012ch} 
that the CS and DS partons can be separated and the CS $\bar{u}^{cs}(x)+\bar{d}^{cs}(x)$ distribution of the proton 
is obtained in the region $0.03 < x < 0.4$ at $Q^2 = 2.5 \,\,{\rm GeV}^2$. This assumes that the distribution of 
$\bar{u}^{ds}(x)+\bar{d}^{ds}(x)$ is proportional to that of $s(x) + \bar{s}(x)$, so that the CS partons can be extracted at
$Q^2 = 2.5\,\, {\rm GeV}^2$ through the relation
\begin{equation}  \label{udCS}
\bar{u}^{cs}(x)+\bar{d}^{cs}(x) = \bar{u}(x)+\bar{d}(x) - \frac{1}{R}(s(x) + \bar{s}(x)),
\end{equation}
where $(s(x) + \bar{s}(x))$ is from the HERMES experiment~\cite{hermes08}, $\bar{u}(x)+\bar{d}(x)$ is from
the CT10 gobal fitting of  PDF~\cite{CT10}, and $R$ is defined as
  \begin{equation}  \label{ratio}
     R=\frac{\langle x\rangle_{s+\bar{s}}}{\langle x\rangle_{u+\bar{u}}(DI)},
   \end{equation}
and we have used the lattice result $R = 0.857(40)$~\cite{doi08} for the extraction. 
\begin{figure}[hbtp]
% \vspace*{1cm}
  \centering
% \hspace{2cm}
  {{\includegraphics[width=0.6\hsize]{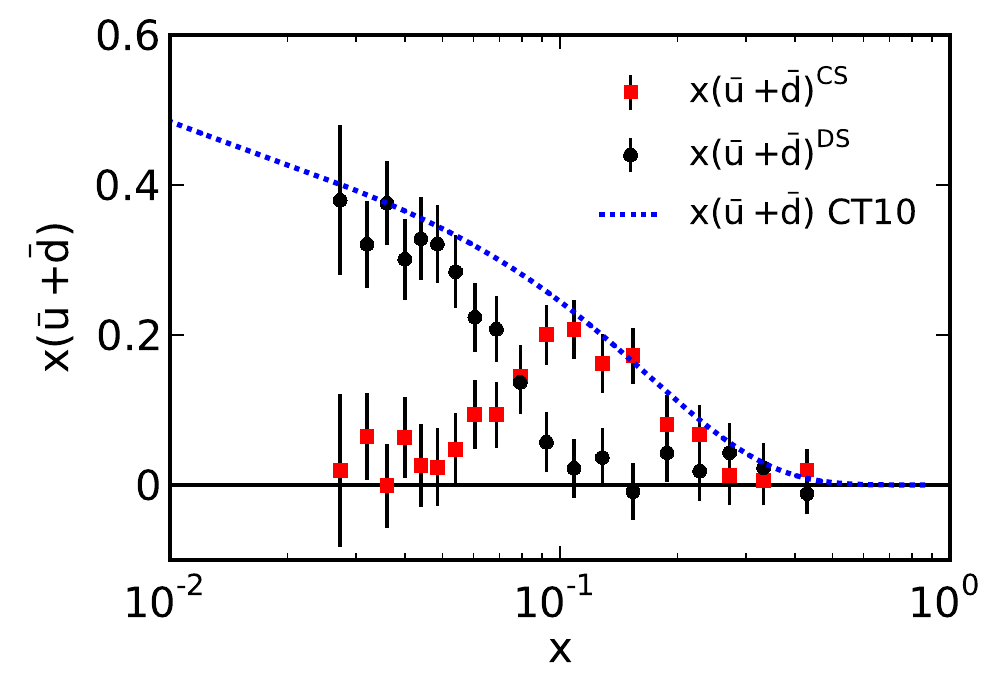}} \label{CSu+d}}
\caption{The $x(\bar{u}^{cs}(x) + \bar{d}^{cs}(x))$ as obtained from Eq.~(2.1)
 is plotted together with  $x(\bar{d}(x) + \bar{u}(x))$ from CT10 and $\frac{1}{R}x(s(x)+\bar{s}(x))$ which
is taken to be $x(u^{ds}(x)+\bar{u}^{ds}(x))$. }
\end{figure}

The results of $x(\bar{u}(x)+\bar{d}(x) - \frac{1}{R}(s(x) + \bar{s}(x))$, 
$x(\bar{u}^{ds}(x)+\bar{d}^{ds}(x) = \frac{1}{R}x(s(x) + \bar{s}(x))$
and \\
\mbox{$x(\bar{u}(x)+\bar{d}(x))$} from CT10 in Fig.~\ref{CSu+d} are plotted in Fig.~\ref{CSu+d} 
to show that the CS and DS have very different $x$-dependence. 

   The extraction of  $\bar{u}^{cs}(x)+\bar{d}^{cs}(x)$ in Eq.~(\ref{udCS}) is based on the assumption
that the distribution of $s(x) + \bar{s}(x)$ is proportional to that of $u^{ds}(x) + \bar{u}^{ds}(x)$ or
$d^{ds}(x) + \bar{d}^{ds}(x)$ so that their ratio can be obtained via the ratio $R$ in Eq.~(\ref{ratio}).
It would be better to calculate  the disconnected insertion of 
$W_{\mu\nu}$ represented in Fig.~\ref{DS} directly on the lattice.

There is a persistent tension between the lattice calculation of
the second moment of $u - d$ and that of the global fitting~\cite{Bali:2014gha,Yang:2015zja}.  
The PDF data is from $\langle x\rangle_{u-d}=\langle x\, (u(x) - d(x))\rangle$ which, according to the path-integral classification, 
corresponds to \mbox{$ \langle x\, (u^{v+cs}(x) - d^{v+cs}(x))\rangle + \langle x\, (\bar{u}^{cs}(x) - \bar{d}^{cs}(x))\rangle$.}
Thus, it is desirable to calculate the $u^{v + cs}(x)(d^{v + cs}(x))$ from Fig.~\ref{val+CS} and  CS in Fig.~\ref{CS} to check the 
Gottfried sum rule violation directly and address the discrepancy of the second moment.

\section{Numeral calculation on the lattice}

     The calculation of Euclidean hadronic tensor $\widetilde{W}_{\mu\nu}(\vec{q}, \vec{p}, \tau)$,  which involves 
 four-point functions, entails two sequential inversions which is somewhat more complicated than the three-point function 
 calculations for  the nucleon form factors and parton moments. Furthermore, the large momentum transfer $|\vec{q}| \gg m_N$
 and large energy transfer $\nu = E_n - E_p$, which are needed to sustain 
 the parton interpretation,  are challenges for the lattice calculation as they will require fine lattice spacings. The small Bjorken 
 $x = \frac{Q^2}{2(E_p \nu - \vec{p}\cdot\vec{q})}$ can be
 accessed with a non-zero nucleon momentum $|\vec{p}| $ which is antiparallel to $\vec{q}$. For example, with $\vec{q} = - \vec{p}$
 ($|\vec{q} = |\vec{p}|=3$ GeV), and $Q^2 = 2\, {\rm GeV^2}$, the Bjorken $x$ is 0.058. As can be seen in Fig.~\ref{CSu+d}, 
 this $x$ should be low enough  to discern the difference between $x(\bar{u}^{cs} + \bar{d}^{cs})$ and $x(\bar{u}^{ds} + \bar{d}^{ds})$. 
 There are applications where lattice calculations require large hadron momenta, such as the semi-leptonic decays of heavy meson 
 $B \rightarrow \pi \ell \bar{\nu}_{\ell}$ and heavy baryon 
 $\Lambda_b \rightarrow p \ell \bar{\nu}_{\ell}$~\cite{Detmold:2015aaa}, and the large momentum
 approach to calculating TMD~\cite{Hagler:2009mb,Musch:2010ka,Engelhardt:2013fra} and PDF~\cite{Ji:2013dva}.
 In view of this, there have been several studies to address this issue on the lattice~\cite{Roberts:2012tp,DellaMorte:2012xc,Bali:2016lva}.
 
  The most challenging and critical task is to convert the $\widetilde{W}_{\mu\nu}$ on the Euclidean lattice in 
  Eq.~(\ref{wmunu_tilde}) to the hadronic tensor $W_{\mu\nu}$ in the Minkowski space. The inverse Laplace transform
  in Eq.~(\ref{wmunu}) is ture only formally but not practical, since there are no lattice data for imaginary $\tau$. Naively,
  one may attempt to consider using $\lim_{\epsilon \rightarrow \infty}\int_{0}^{\tau_{max}}d\tau e^{(\nu+ i \epsilon)\tau} 
  \widetilde{W}_{\mu\nu}(\vec{q}^{\,2}, \tau)$ to pick out the delta function $\delta (v - (E_n - E_p)) $ to complete the
  4-D delta function in Eq.~(\ref{w}). However, this introduces an exponentially function 
  $e^{(\nu - E_m + E_p)\tau}= e^{(E_n - E_m)\tau}$ in the integrand for $E_m \neq E_n$. For those intermediate states lower than 
  $E_n$, i.e. $E_m < E_n$,  they will have exponentially increasing contributions compared to the constant $\delta$-function that 
  one is trying to pull out~\cite{xu15}. This will not work for DIS where the virtuality is large, i.e. $\nu \gg m_N$. 
  
   $W_{\mu\nu} (q^2, \nu)$ is just the spectral density $\rho(\omega)$ with the excitation energy $\omega = \nu$ .  
   The task is to solve the inverse problem in order to find  the spectral density $\rho(\omega)$ from its spectral representation in the integral
  \begin{equation}
  D(\tau) = \int K(\tau, \omega) \,\rho(\omega)\, d \omega
  \end{equation}
 where $D(\tau) = \widetilde{W}_{\mu\nu}(\tau)$ and the kernel $K(\tau, \omega) = e^{- \omega\, \tau}$.
 A common approach to this inverse problem is the Maximum Entropy Method (MEM)~\cite{bryan90}.
 Bayes' theorem \mbox{$P[\rho|D, I] = \frac{P[D|\rho, I]\, P[\rho|I]}{P[D|I]}$}, where 
  $P[D|\rho, I] \propto e^{-L}$ is the likelihood probability with $L = \frac{1}{2} \chi^2$ being the $\chi^2$ for
  the theoretical model with the discrete version of the spectral representation 
  $D_{\rho} (\tau_i) = \sum_{\ell=1}^{N_{\omega}} e^{- \omega_{\ell}\, \tau_i}\, \rho_{\ell}\,\, \Delta \omega_{\ell}$.
  The standard MEM is to find $\rho_{\ell}$ to maximize $P[\rho|D, I] \propto e^{\,\alpha S - L}$ where $S$ is the 
  Shannon-Jaynes information entropy. Recently, there is an improved MEM which can lead to more stable fit~\cite{Burnier:2013nla}.
  The improved probability to maximize is 
  \begin{equation}
  P[\rho|D, I] \propto e^{\,\alpha S - L- \gamma(L - N_{\tau})^2} ,
  \end{equation}  
where $N_{\tau}$ is the number of $\tau_i$ and $S$ is modified to 
$S = \int d\omega \lbrack 1 - \rho(\omega)/m(\omega) - \ln (\rho(\omega)/m(\omega))\rbrack$.
Given this new promising approach to the inverse problem, it is worthwhile to tackle the hadronic tensor  $W_{\mu\nu}$ 
calculation on the lattice. It does require precise data for the improved MEM, however.

\section{Comparison with the large momentum approach to calculating PDF}
  
  There is another approach to calculating PDF on lattice proposed by Xiangdong Ji~\cite{Ji:2013dva}. The idea is to 
  calculate the `quasiparton' $\tilde{q}(x, \Lambda, P_z)$ on the lattice at a finite $P_z$ and a UV scale of $\Lambda$ and
  match it to the parton distribution $q(x, \mu)$ perturbatively where $\Lambda$  will turn into the renormalization scale $\mu$
  for the parton distribution at the infinite momentum frame (i.e. $P_z \rightarrow \infty$). Besides the perturbative matching
  from the quasiparton to the parton, this approach needs a renormalization procedure for the non-local operator which involves 
 a spatial Wislon line to match the lattice results to those in the continuum. Furthermore, the hadron momentum $P_z$ needs to be
 large. From the lattice calculation of  $\tilde{q}(x, \Lambda, P_z)$ with $P_z \sim 1$ GeV~\cite{Lin:2014zya,Alexandrou:2015rja}, 
 it is found that the antiparton
 which is obtained from the negative $x$ region, i.e. $\bar{q}(x) = - q(-x)$ is negative. This unphysical result is possibly
 due to the fact that $P_z \sim 1$ GeV is not large enough. This feature can be used to test how large a $P_z$ is needed
 to have a positive probability for the CS part of the antipartons.   
 
 The present hadronic tensor approach requires calculation of 4-point functions  on the lattice which is more numerically intensive
 than the large momentum approach which entails 3-point functions. Yet, it is simpler theoretically. There is no renormalization
 for the vector currents (other than the normalization to satisfy Ward identity for the non-conserved current on the lattice).
 Since $W_{\mu\nu}$ is Lorentz covariant, there is no need to work in the large momentum frame as in the large momentum
 approach. The frame independent $W_1$ and $W_2$ in $W_{\mu\nu}$ can be obtained with any hadron momentum.
 The most challenging part is to convert the Euclidean $\widetilde{W}_{\mu\nu}(\vec{q},\vec{p},\tau)$ to $W_{\mu\nu}(q^2, \nu)$
 in Minkowski space. The improved MEM will be adopted to see how successful it is in this regard. Once the $W_{\mu\nu}(q^2, \nu)$
 is obtained at the continuum and large volume limits and at the physical pion point, one can apply QCD factorization to fit the PDF to
 NNLO as is currently done in global fitting of the experimental lepton-hadron cross-sections. The advantage of  having
 lattice results to supplement experimental data is that the flavors and the partons degrees of freedom, i.e. the valence, the CS and DS represented
 in Figs.~\ref{val+CS}, \ref{CS}, and \ref{DS} are readily separated to better understand the role of each degree of freedom, their
 evolutions, as well as the flavor dependence. 

 This work is partially supported by DOE grant no.  DE-SC0013065.

\end{document}